\documentclass[12pt,a4paper]{article}

\begin{document}

\title{Pseudo-rapidity Distributions
of Charged Hadrons in $pp$ and $pA$ Collisions at the LHC\thanks
{Supported by Natural Science Foundation of Hebei Province
(A2012210043)}} \small{\date{}}
\author{\small {WANG Hong-Min$^{1,2}$\thanks{E-mail:whmw@sina.com.cn},
  LIU Jia-Fu$^{1}$, HOU Zhao-Yu$^{3}$, SUN Xian-Jing$^{4}$}\\
$^{1}$\small{Physics Department, Academy of Armored Forces
Engineering of PLA,
Beijing 100072, China}\\
$^{2}$\small{Physics Department, Tsinghua University,
Beijing, 100084, China}\\
$^{3}$\small{Physics Graduate School, Shijiazhuang Railway
Institute, 050043, China}\\
$^{4}$\small{Institute of High Energy Physics, Chinese Academy of
Sciences, Beijing 100049, China}}
 \maketitle
{\textbf{Abstract:} In the framework of Color Glass Condensate, the
pseudo-rapidity distributions of charged hadrons in $pp$ and $pA$
collisions at the LHC are studied with the UGD function from the GBW
model. With a $\chi^{2}$ analysis of the CMS data in $pp$ collisions
at $\sqrt{s}=0.9,~2.36,~7$ TeV, the normalization factor is obtained
and the theoretical results are in good agreement with the
experimental data. Then, considering the influence of nucleon hard
partons distribution on the number of participants in $pA$
collisions by a Glauber Monte Carlo method, we give the predictive
results for the multiplicity distributions in $p+Pb$ collisions at
$\sqrt{s}=4.4$ TeV.
}\\



{\textbf{1 Introduction}}
\\

At super high energies, perturbative Quantum Chromodynamics (pQCD)
predicts that the small-$x$ gluons in a hadron wave function should
form a Color Glass Condensate (CGC) [1], which is characterized by
strong coherent gluon fields leading to parton saturation. After
that, signals of parton saturation have already been observed both
in electron-proton ($ep$) deep inelastic scattering at HERA [2] and
in deuteron-gold collisions at Relativistic Heavy-Ion Collisions
(RHIC) [3,4]. However, the research of the nature of CGC need more
confirmation, so it is still an active subject in both theoretical
and experimental sides. Recently, the data for charge hadron
multiplicities in $pp$ collisions are probed by Large Hadron
Collider (LHC) [5,6], and soon the first data on $p+Pb$ collisions
at $\sqrt{s}=4.4$ TeV will also be given. The data at the LHC will
allow to probe the nuclear gluon distributions at very small Bjorken
$x$ ($\sim 10^{-6}$). Thus, these measurements are very important
for testing the nature of the CGC. In this paper, basing on the CGC
formalism, we will investigate the charge hadron multiplicity
distributions in $pp$ collisions and give the predictive results for
$p+Pb$ collisions.

In the CGC formalism, the cross section for inclusive hadron
production can be given by the convolution of the unintegrated gluon
distribution (UGD) of the proton
 (or nucleus) from the projectile and the target [1,7]. For the
 charge hadron pseudo-rapidity distributions, the cross section can be
 obtained by integrating the inclusive production over $p_{t}$ and a
 Jacobian transformation [7,8].
  In this paper, the simple UGD function from the Golec-Biernat
and W$\mathrm{\ddot{u}}$sthoff (GBW) model [9], which has
successfully described both the HERA and RHIC data, is used for $pp$
considered. Through a $\chi^{2}-$analysis of the CMS data
  for $pp$ collisions, the normalization factor
  $K$ that describes the conversion of partons to hadrons can be obtained.

In $pA$ collisions, the number of participating nucleons in the
collisions must be considered. The simple and appropriate method to
calculate the number of participants is the Glauber Monte-Carlo
(GMC) approach [10]. At super high energy domain, the contribution
from small-$x$ gluons dominate the mechanism of inelastic hadronic
collisions and the influence of the transverse spatial distribution
of hard partons in the nucleon must be considered in the GMC
approach [11-13]. The simply and popularly used method to consider
the hard partons transverse distribution is considered the nucleon
as a hard sphere. In this paper, in order to give a more
 accurate predictive results for $pA$ collisions,
 the transverse distribution derived from the
$J/\Psi$ photo-production data in the nucleon is used [11].
\\

{\textbf{2 Method}}
\\

In the CGC formalism [1,7], the formula for the inclusive production
can be given by
\begin{equation}
E\frac{d\sigma}{d^{3}p}=K \frac{2}{C_{F}}
\frac{1}{p^{2}_{t}}\int^{p_{t}}dk^{2}_{t}\alpha_{s}\varphi_{p_{1}}
(x_{1},k_{t}^{2})\varphi_{p_{2}}(x_{2},(p-k)_{t}^{2}),
\end{equation}
where $C_{F}=(N_{c}^{2}-1)/(2\pi N_{c})$,
$x_{1,2}=(p_{t}/\sqrt{s})\mathrm{exp}(\pm y)$ and $\sqrt{s}$ is the
center of mass energy. $\varphi_{p}$ is the unintegrated gluon
distribution of a proton. The normalization factor $K$ can be
determined by a global fit to $pp$ data at various energies.

 The multiplicity distribution per unit rapidity can be given by
 integrating Eq.(1) over $p_{t}$
$$\frac{dN}{dy}=\frac{1}{S}\int d^{2}p_{t}E\frac{d\sigma}{d^{3}p},$$
where $S$ is either the inelastic cross section for the minimum bias
multiplicity, or a fraction of it corresponding to a specific
centrality cut.
 For the main contribution to
Eq. (1) is given by two regions of  integration over $k_{t}$:
$k_{t}\ll p_{t}$ and $|\vec{p}_{t}-\vec{k}_{t}|\ll p_{t}$, Eq. (1)
can be rewritten as
\begin{equation}
\frac{dN}{dy}=\frac{K}{S}\frac{2\alpha_{S}}{C_{F}}\int
\frac{dp_{t}^{2}}{p_{t}^{2}}
[\varphi_{p_{1}}(x_{1},p_{t}^{2})\int^{p_{t}}dk_{t}^{2}\varphi_{p_{2}}(x_{2},k_{t}^{2})+
\varphi_{p_{2}}(x_{2},p_{t}^{2})\int^{p_{t}}dk_{t}^{2}\varphi_{p_{1}}(x_{1},k_{t}^{2})].
\end{equation}
Here, for the unintegrated gluon distribution, $\varphi_{p}$, the
one from the GBW model will be used [9]
\begin{equation}
\varphi_{p}(x,p^{2}_{t})=\frac{3\sigma_{0}}{4\pi^{2}\alpha_{s}(Q_{s})}\frac{p_{t}^{2}}{Q_{s}^{2}(x)}
\mathrm{exp}(-\frac{p_{t}^{2}}{Q_{s}^{2}(x)}),
\end{equation}
where the saturation scale is taken as [14]
\begin{equation}
Q_{s,p}^{2}(y)=Q_{0}^{2}(x_{0}\frac{\sqrt{s}}{Q_{0}}\mathrm{exp}(\pm
y))^{\bar{\lambda}},
\end{equation}
with the parameters $Q_{0}=0.6$ GeV, $x_{0}=0.01$ and
$\bar{\lambda}=0.205$. The running coupling constant $\alpha_{s}$,
is assumed to freeze at $\alpha_{\mathrm{max}}=0.52$ [15]
\begin{equation}
\alpha_{s}(Q^{2})=\mathrm{min}[\frac{12\pi}{27\mathrm{log}\frac{Q^{2}}{\Lambda^{2}}},\alpha_{\mathrm{max}}],
\end{equation}
where $\Lambda=0.226$. In order to account for large-$x$ effects in
the gluon distribution, the distribution function is always
multiplied by $(1-x)^{4}$.

To calculate the distribution verse pseudo-rapidity, one should
express rapidity $y$ in terms of pseudo-rapidity $\eta$
\begin{equation}
y(\eta)=\frac{1}{2}\mathrm{log}\frac{\sqrt{\mathrm{cosh}^{2}\eta+\mu^{2}}+\mathrm{sinh}\eta}
{\sqrt{\mathrm{sinh}^{2}\eta+\mu^{2}}+\mathrm{cosh}\eta},
\end{equation}
and the Jacobian can be obtained by
\begin{equation}
J(\eta)=\frac{\partial y}{\partial \eta}=\frac{\mathrm{cosh}
\eta}{\sqrt{\mathrm{cosh}^{2}\eta+\mu^{2}}},
\end{equation}
where the scale $\mu(\sqrt{s})=0.24/(0.13+0.32\sqrt{s}^{0.115})$
with $\sqrt{s}$ expressed in units of TeV [14].

For $pA$ collisions, the saturation scale of the nucleus can be
given as
\begin{equation}
Q_{s,A}^{2}(y)=N_{\mathrm{part}}Q_{s,p}^{2}(y),
\end{equation}
where $N_{\mathrm{part}}$ is the number of participating nucleons in
the collisions. In the  GMC approach, the number of participants can
be given by [10]
\begin{equation}
N_{\mathrm{part,A}}(\vec{b})=\sum_{i=1,2...A}P(|\vec{b}-\vec{r}_{i}|),
\end{equation}
where $\vec{b}$ is the impact parameter of the $pA$ collisions, and
the set ${\vec{r}_{i}}$, which corresponds to the coordinates of the
nucleons in the target, can be picked randomly according to a
Woods-Saxon distribution. In the GMC framework, the nucleons is
always simply considered as "hard sphere" (HS),  and the function
\begin{equation}
P^{\mathrm{HS}}(|\vec{b}-\vec{r}_{i}|)=\Theta(|\vec{b}-\vec{r}_{i}|-d_{\mathrm{max}}),
\end{equation}
where $d_{\mathrm{max}}=\sqrt{\frac{\sigma_{in}(\sqrt{s})}{\pi}}$
with $\sigma_{in}(\sqrt{s})=52,60,65.75,70.45$ mb at
$\sqrt{s}=0.9,2.36,4.4,7$ TeV, respectively. Here, the nucleon
partons transverse distribution derived from the $J/\psi$
photo-production data is used, and this transverse distribution can
be described by a dipole form [13]
\begin{equation}
P^{\mathrm{D}}(|\vec{b}-\vec{r}_{i}|)=m_{g}^{2}/(4\pi)(m_{g}|\vec{b}-\vec{r}_{i}|)K_{1}(m_{g}|\vec{b}-\vec{r}_{i}|),
\end{equation}
 where $K_{1}$ denotes the modified Bessel function and the mass parameter $m_{g}\sim1.1$ GeV$^{2}$.
The ratio of $N_{\mathrm{part}}$ with the dipole model to that with
the hard sphere model is shown in Fig. 1. Here, the woods saxon form
is used for the nucleon distribution in plumbum (Pb) [16]. It is
shown that a clearly downward trend can be seen in the domain $b>6$
fm.
\\

\textbf{3 Results and Discussion}
\\

In order to obtain the normalization factor from the data in $pp$
collisions, we must introduce the $\chi^{2}-$analysis method [17]
\begin{equation}
\chi^{2}=\sum_{j}^{n}\frac{(\frac{dN}{d\eta}|^{\mathrm{data}}_{\mathrm{pp},j}-
\frac{dN}{d\eta}|^{\mathrm{theo}}_{\mathrm{pp},j})^{2}}{(\frac{dN}{d\eta}|^{\mathrm{err}}_{\mathrm{pp},j})^{2}},
\end{equation}
where $\frac{dN}{d\eta}|^{\mathrm{data}}_{\mathrm{pp},j}$
($\frac{dN}{d\eta}|^{\mathrm{theo}}_{\mathrm{pp},j}$) indicates the
experimental data (theoretical values) for the charge hadron
pseudo-rapidity distributions in $pp$ collisions, and
$\frac{dN}{d\eta}|^{\mathrm{err}}_{\mathrm{pp},j}$ denotes the
systematic errors in the experiment. With
$\chi^{2}_{\mathrm{min}}/$(degree of freedom)=0.2597, we find the
normalization factor $k(=K/S_{\mathrm{pp}})$ is equal to 0.098.  The
theoretical results in $pp$ collisions at
$\sqrt{s}=0.9(\mathrm{a}),~2.36(\mathrm{b}),~7(\mathrm{c})$ TeV are
shown in Fig. 2. As a contrast, the results with the KLN model is
also given, and the solid and dashed curves are the results with the
GBW model and KLN model, respectively. It is shown that the
theoretical results with both of them are in good agreement with the
experimental data. Fig.2 also shows that the theoretical results
with the GBW model are lower than that with the KLN model at lager
$\eta$, these should be verified for the further experiments.

For $pA$ collisions, the average hadron multiplicities  can be
obtained by
\begin{equation}
\frac{dN}{d\eta}|_{\mathrm{pA}}=\frac{\int_{b_{1}}^{b_{2}}db 2\pi b
\frac{dN}{d\eta}|_{\mathrm{pA}}(b)}{\int_{b_{1}}^{b_{2}}db 2\pi b
\{1-[1-\sigma_{in}^{pp}\hat{T}_{A}(b)]^{A}\}},
\end{equation}
 where $\hat{T}(b)$ is the thickness function.
 Fig. 3 show the predictive results for $p+Pb$ collisions at
$\sqrt{s}=$4.4 TeV for different centrality: 0-100\% (solid curve),
0-50\% (dashed curve), 50\%-100\% (dotted curve). The results will
be checked by the future experiments.

In summary, according to the CGC formalism, we have calculated the
hadron pseudo-rapidity distribution at the LHC. With the
$\chi^{2}-$analysis of the experimental data in $pp$ collisions, the
normalization factor is obtained and the theoretical results are in
good agreement with the experimental data. In the end of this paper,
the predictive results at different centrality are also given, and
the theoretical results will be validated by the future experiments.

\begin{newpage}

\end{newpage}

\begin{newpage}

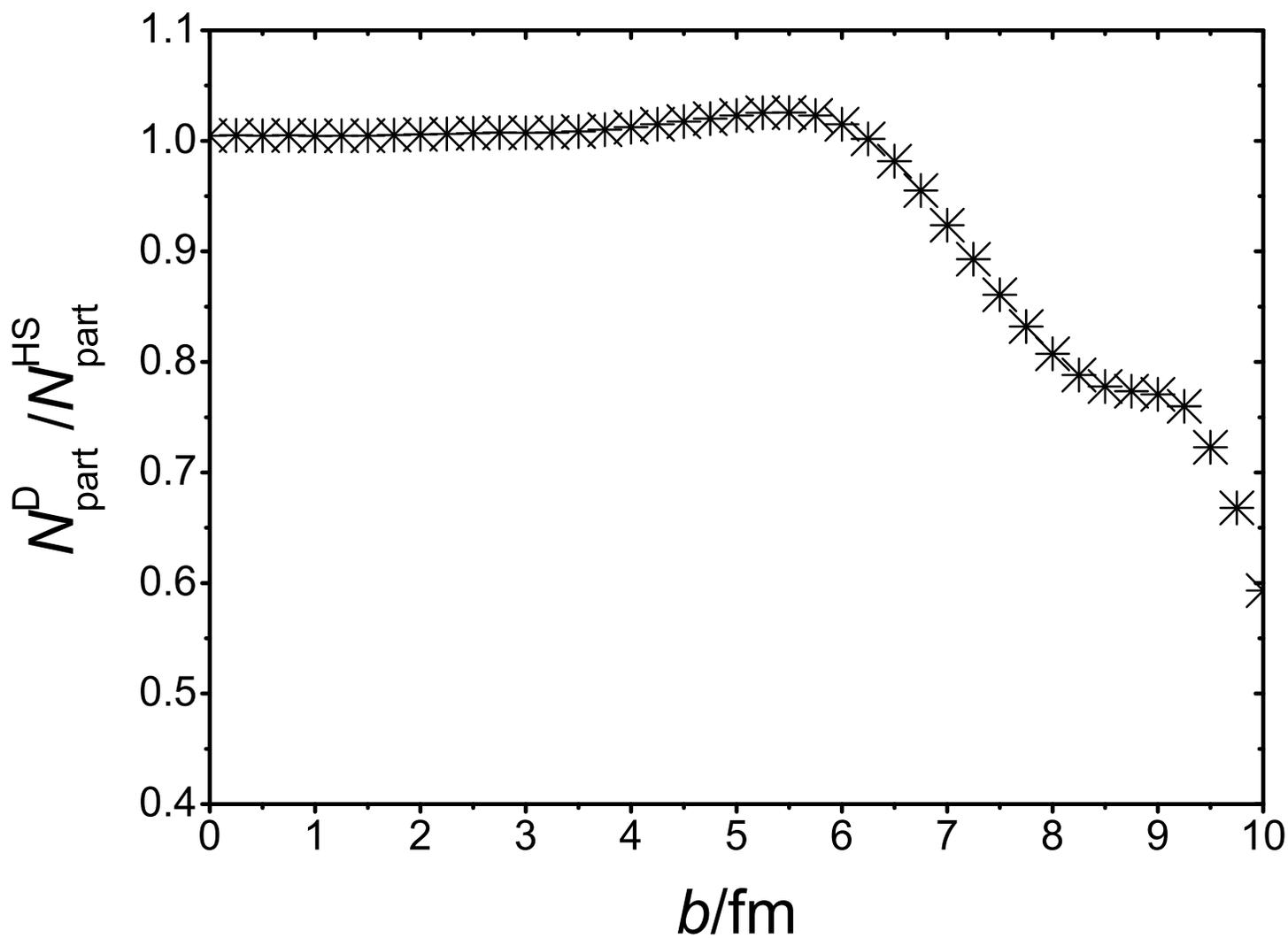
\begin{figure}
\centering
{figure1} \caption{ The ratio of $N_{\mathrm{part}}$ with
the dipole model to that with the hard sphere model.}
\end{figure}

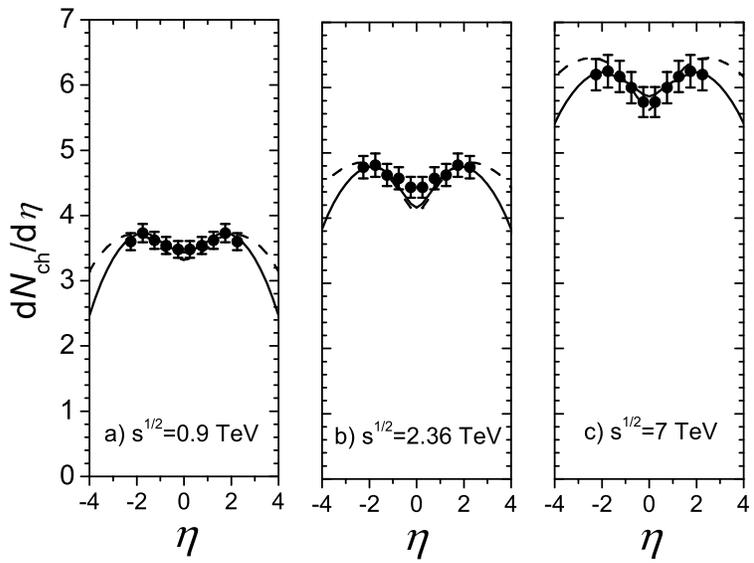
\begin{figure}
\centering
{figure2} \caption{Pseudo-rapidity distribution of
charged hadrons in $pp$ collisions at $\sqrt{s}=0.9$ TeV (a), 2.36
TeV (b) and 7 TeV (c). The solid and dashed curves are the results
of the GBW model and the KLN model, respectively. The data come from
CMS.}
\end{figure}

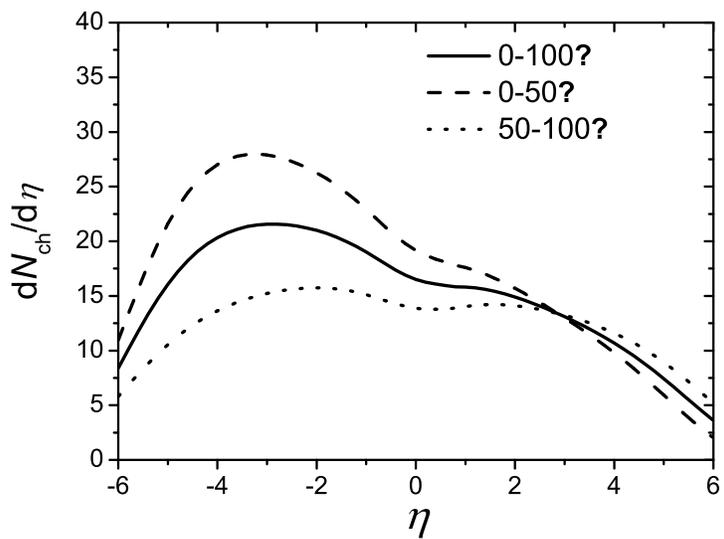
\begin{figure}
\centering
{figure3} \caption{Pseudo-rapidity distribution of
charged particles in minimum bias $p+Pb$ collisions at $W=4.4$ TeV
with the GBW model for different centrality cuts.}
\end{figure}

\end{newpage}
\end{document}